# Stacking fault energy prediction for austenitic steels: thermodynamic modeling vs. machine learning

Xin Wang, Wei Xiong *

*Physical Metallurgy and Materials Design Laboratory,*
*Department of Mechanical Engineering and Materials Science,*
*University of Pittsburgh, Pittsburgh, PA 15261, USA*
* Corresponding author: w-xiong@outlook.com, weixiong@pitt.edu
Tel.: +1-(412) 383-8092, Fax: +1-(412) 624-4846
URL: http://www.pitt.edu/~weixiong

**Abstract**

Stacking fault energy (SFE) is of the most critical microstructure attribute for controlling the deformation mechanism and optimizing mechanical properties of austenitic steels, while there are no accurate and straightforward computational tools for modeling it. In this work, we applied both thermodynamic modeling and machine learning to predict the stacking fault energy (SFE) for more than 300 austenitic steels. The comparison indicates a high need of improving low-temperature CALPHAD (CALculation of PHAse Diagrams) databases and interfacial energy prediction to enhance thermodynamic model reliability. The ensembled machine learning algorithms provide a more reliable prediction compared with thermodynamic and empirical models. Based on the statistical analysis of experimental results, only Ni and Fe have a moderate monotonic influence on SFE, while many other elements exhibit a complex effect that their influence on SFE may change with the alloy composition.

## 1. Introduction

Alloy with high strength and excellent ductility is one of the ultimate goals for materials design. A practical pathway is to introduce the secondary deformation mechanisms, such as twinning-induced plasticity (TWIP) and transformation-induced plasticity (TRIP), which can improve the strength and ductility simultaneously during deformation [1,2].

Stacking fault energy (SFE) is the energy that related to dissociating a perfect dislocation into two partial dislocations along with the formation of a stacking fault [3]. The SFE is a useful indicator to predict which secondary deformation mechanism operates. Generally, TRIP activates when SFE is lower than 20 mJ/m$^2$ [4], and TWIP is achievable if the SFE lies between 20 – 40 mJ/m$^2$ [5]. Therefore, various efforts have been made to the SFE determination. Experimental methods, including transmission electron microscopy [6–8] and X-Ray/Neutron diffraction [9,10], are time-consuming and complex, which delays the process of new materials discovery. Computational methods such as empirical equations, *ab initio* calculations, and thermodynamic models serve as an alternative solution. Although various empirical equations have been proposed [11–13], most of them are localized for a confined composition. Recently, de Bellefon *et al*. [14] collected 144 data points and successfully predicted SFE using a linear regression method. However, the prediction model was established for stainless steel, and there is a need to build a model suitable for other austenitic steel, such as high Mn steels, with higher accuracy by utilizing a more comprehensive database and applying different algorithms. A thermodynamic model was proposed by Olson and Cohen [15] and has been adapted in many modeling of SFE in the austenitic steels [16–19]. Within this approach, an intrinsic stacking fault is defined as an hcp phase with two boundaries shared with the fcc matrix. The reliability of thermodynamic models heavily depends on the quality of CALPHAD (CALculation of PHAse Diagrams) databases. However, in most of the past research, the SFE model and Gibbs free energy functions for phases are designed for steels with 2-3 alloying elements [16–19], while modeling the multicomponent systems is challenging since it involves many parameters. Moreover, they have not been verified for alloys with a wide composition range [16,19,20]. Last but not least, it is challenging and time-consuming for *ab initio* methods [21,22] to deal with chemical and magnetic energy contributions in complex multicomponent alloys [23]. Certain works underestimate the SFE and even report negative values [24,25]. Thus, some models are often only capable of predicting trends due to simple alloying effects with a limited composition range. A promising way to leverage the wealth of data and circumvent the difficulty of SFE prediction is by applying data-driven methods [26,27]. Machine learning (ML) is useful in extracting knowledge from multi-dimensional data and modeling the relationships between the targeted property and its related features [28].

      In recent years, only a few studies have applied ML for SFE prediction [29,30]. Das [29] predicted SFE using an artificial neural network (ANN) with 100 compositions as an

input. But, that work did not incorporate temperature into the model while the temperature is an important factor controls SFE [31], and ANN requires an extensive database for generating a reliable model, which may limit the accuracy of this work. Chaudhary *et al.* [30] built a classifier that categorizes compositions into high, medium, and low SFE. However, the prediction of the actual SFE value is crucial, since the SFE value is a key parameter in modeling the critical stress for twinning and the mechanical properties [32,33]. Thus, a systematic study for understanding the relationship between composition and SFE, together with building an accurate SFE predictor, is imperative. Overall, this work (i) assessed the quality of the CALPHAD-based thermodynamic model-prediction, and revealed the importance of robust CALPHAD database on accurate SFE prediction; (ii) discussed the influence of alloying elements on SFE through a statistical approach and found Ni and Fe have a moderate monotonic influence on SFE while other elements might have a complex effect; and (iii) predicted SFE using ML, and proved the performance of the ML model developed in this work was superior to the thermodynamic and empirical models.

## 2. Methods

Figure 1 depicts the framework used in this work. We firstly performed a comprehensive literature survey and constructed an experimental database containing 349 entries with temperature, composition, and SFE [6,7,34–43,8,44–53,9,54–63,10,64–73,11,74,12,13,15,19] from those references without limitation on the SFE measurement techniques, sample preparation methods, etc.. But compositions containing uncommon elements such as W, V were not collected since the number of data is limited. Further, we randomly split them into train and test datasets. Our dataset covered a broad range of compositions, and its descriptive statistics are listed in Table 1. Secondly, all entries in this database were screened by the thermodynamic model to calculate the SFE using Eq. (1) [15,19]:

$$\gamma = 2\rho \Delta G^{fcc \to hcp} + 2\sigma^{fcc/hcp}, \quad \rho = \frac{4}{\sqrt{3}a_{fcc}^2}\frac{1}{N_A} \quad (1)$$

where $\gamma$ is the SFE (mJ/m$^2$), $\Delta G^{fcc \to hcp}$ is the Gibbs energies difference between the fcc and hcp phases [75]. Thermo-Calc [76] software with TCFE9 and TCHEA3 databases were used to calculate the Gibbs energy. The TCFE9 is mainly designed for steels, and TCHEA3 is constructed for multi-principal element alloys. These commercial databases developed by experts are expected to achieve reasonable thermodynamic prediction over a wide composition range for multicomponent alloys by comparing with the ones reported in the literature, which are usually designated to specific steels with limited composition and temperature ranges. $\rho$ is the molar surface density (mol/m$^2$) of {111} plane, $N_A$ is the Avogadro's number, $a_{fcc}$ = 0.36 nm is the lattice parameter of fcc, $\sigma^{fcc/hcp}$ = 8 mJ/m$^2$ is interfacial energy [75]. Furthermore, we performed Spearman's correlation analysis using Python [77] and SciPy [78] to find the influence of alloying elements on SFE. Three different sets of features named as Standard, WithTCFE9, and WithTCHEA3 were built to find out whether the thermodynamic model can enhance the model predictability. An appropriate selection of features, which distinguish the material and describe the property of interest, can lead to better performance of the ML model. Further, we evaluated the performance of 19 algorithms available in Scikit-learn [79] with different hyper-parameters for the three different feature sets using the 10-fold cross-validation [80] to discover the model with highest accuracy and generalizability. The 75% train data is randomly split into ten subsets, and the model is fitted with nine subgroups and tested with the remaining subgroup [81]. After training and testing for 10 times, the average values of metrics such as the root mean square error (RMSE) and mean absolute error (MAE) were calculated using Eqs. (2) and (3), respectively:

$$RMSE = \sqrt{\frac{\sum_{i=1}^{n}(SFE_i^E - SFE_i^P)^2}{n}} \quad (2)$$

$$MAE = \frac{\sum_{i=1}^{n}|SFE_i^E - SFE_i^P|}{n} \quad (3)$$

where $n$ is the number of data points in the evaluation, $SFE_i^E$ and $SFE_i^P$ are experimental and predicted SFE of the datapoint $i$, respectively. Finally, the model with the lowest RMSE was selected and compared with the empirical and thermodynamic models.

## 3. Results and discussion

### 3.1 Evaluation of the thermodynamic model

Evaluation of SFE prediction by the thermodynamic model is presented in Fig. 2. Although, a few data points lie in the black dash line in Figs. 2(a) and 2(b), indicating the equivalency between the thermodynamic model prediction (SFE$_{calc}$) and experiments (SFE$_{exp}$), but many other data points show a large deviation. Moreover, SFE$_{exp}$ varies from 3 to 80 mJ/m$^2$, while SFE$_{calc}$ with TCFE9 varies between -100 to 150 mJ/m$^2$ and the TCHEA3 predicted values were as high as 800 mJ/m$^2$. The large discrepancy between SFE$_{calc}$ and SFE$_{exp}$ may originate from the following reasons. First, the interfacial energy used in this work and other reports is a constant, while it is a composition-dependent variable. Also, the difference in interfacial energy among previous studies differs by more than 20 mJ/m$^2$, which could introduce a large uncertainty [1,18,20,82–84]. Secondly, most SFE measurements are performed at room temperature while the low-temperature CALPHAD databases for multicomponent systems lack precision, and the current works mainly focus on the pure element and binary reassessment [85,86]. In order to show the effect of temperature on the thermodynamic model accuracy, the whole dataset was split into three different groups: elevated temperature (300 < T < 600K, 12 samples), room temperature (300 K, 290 samples), and low temperature (94 < T < 300 K, 47 samples). The MAE of the CALPHAD prediction with different databases was calculated and listed in Table 2. The prediction error is relatively small at the elevated temperature compared with room temperature and low temperature. Moreover, Spearman's correlation analysis of the temperature and the absolute error of CALPHAD-based calculation was performed. For TCHEA3 and TCFE9 databases, Spearman's correlation coefficient is -0.24 and -0.19, respectively. The Spearman's rank correlation coefficient *r* is a statistic measure of the strength of the monotonic relationship between two data, where *r* lies between -1 to 1. A positive value corresponds to a positive monotonic relation, i.e., as one variable increases, another variable also increases. A negative value indicates a negative monotonic relationship that when one variable increases, another variable will decrease. And 0 denotes that the two variables are not monotonically related. Suppose the absolute value of *r* is <0.4, 0.4-0.6, >0.6, the correlation between two variables can be interpreted as weak, moderate, and strong, respectively [87]. According to the negative correlation coefficient *r*, it is clear that when the temperature is increasing, the error of CALPHAD simulation has a weak monotonic trend to decrease. Additionally, the magnetic contribution to SFE is significant at low temperatures, and there is a need for establishing a robust and sophisticated magnetic model to improve the accuracy of the thermodynamic model [88,89]. Though the thermodynamic model is not accurate for all steels tested in this work, it is useful for certain alloy systems.

According to Fig. 2(c), the two databases generate a good prediction for several alloy systems. For TCFE9, the error in Fe-Mn-Si-Al, Fe-Mn-Si, Fe-Mn, and Fe-Mn-Al systems are relatively small. For TCHEA3, the error in Fe-Cr-Ni, Fe-Cr-Mo-Ni, Fe-Mn-Al, and Fe-Mn systems is acceptable. Spearman's rank correlation coefficients (*r*) for SFE$_{calc}$ and SFE$_{exp}$ are shown in Fig. 2(d). For Fe-Mn-Si-Al, Fe-Mn-Al, and Fe-Cr-Mn-Ni systems, the *r* values between $SFE_{exp}$ and $SFE_{calc}$ using TCFE9 are higher than 0.5, which indicates TCFE9 can predict the trend of SFE change with different elements in steels containing Mn. However, for the Fe-Cr-Ni system, TCHEA3 performs better with *r* value around 0.75, indicating that it is suitable for steels with high Cr and Ni. In summary, the performance of the thermodynamic model heavily depends on the quality of CALPHAD databases, which should be carefully chosen depending on the alloy composition.

*3.2 Alloying effects on SFE*

Understanding the influence of alloying elements on SFE is crucial for alloy design and has attracted various studies [6,25,90]. However, due to the limited time and resources, previous work only focused on limited alloys to draw a conclusion. Vitos *et al.* [21] pointed out that the alloying effect on SFE is a function of both alloying element content and the host composition. Thus, the general influence of elements on SFE remains unclear. Here, we calculated the Spearman's *r* and the *p*-value for interpreting the relationship between each element and the SFE without considering the host composition. A *p*-value serves as evidence against a null hypothesis, and a smaller *p*-value indicates a low possibility that the null hypothesis is true. In this work, the null hypothesis is that the two variables are uncorrelated but still generate the same Spearman's rank correlation. If a *p*-value is 0.05 (criteria used in Fig. 3), there is only a 5% chance to get the same correlation coefficient *r* for two unrelated variables [91]. According to Fig. 3, Ni has the most pronounced effect in increasing the SFE regardless of the host composition, which agrees with the analysis by Das [29]. It was reported that from a total of 20 reports, 17 indicated an increasing effect. C and Mn do not show a strong effect of increasing/decreasing the SFE based on our study. Because the *r* value is close to 0 and the p-value is larger than 0.05, which confirms that C and Mn do not have a statistically significant monotonic relationship with SFE. The importance of features in the GB has also been calculated and shown in Fig. 3(b). A high value denotes more times that this feature has been used as a critical decision in the model, and the feature is important in promoting the performance of the model [92]. Based on Fig. 3(b), C, and Mn are important for the SFE predictor generated through the gradient boosting (GB) algorithm [93]. This is because the effect of these elements on SFE is more complicated than that of Ni and varies with the host alloy, which has also been verified by the literature review [29]. For example, an *ab initio* study found that the addition of Mn into Fe-Cr-Ni stainless steel will always decrease the SFE at 0K, and only increase the SFE at room temperature and when Ni content is higher than 16 % [94]. Meanwhile, Pierce et al. [19] showed that when Mn content was increased from 22 wt.% to 28 wt.% in Fe-xMn-3Al-3Si steel, the SFE increased monotonously. However, the Si [8,45] is generally considered to decrease the SFE, we get a contradictory conclusion, which may be attributed to the collective effect from the data pertaining to different alloy systems.

### *3.3 Machine learning model for SFE prediction*

Table 3 summarizes the mean RMSE and MAE of the 10-fold cross-validation for each model with optimized hyper-parameters. We found that the ensembled tree algorithms, including random forest [95], GB, and XGBoost [96] algorithms, generate a more accurate model than other algorithms tested in this work. The GB model has MAE around 5.5 mJ/m$^2$ for all input sets. The RMSE is sensitive to large prediction errors and is reported to be near 8 mJ/m$^2$. This implies that few large deviations happen during the prediction. And, the overall performance of the ML models is reasonable for such an error. In our study, the ensembled methods always perform better than the multilayer perception (MLP, an ANN model) based on different metrics, suggesting that there are better algorithms for predicting SFE than ANN as stated in the introduction. We also found that the overall performance is similar among the three input sets, as indicated in Fig. 1. Although the performance of adding TCFE9 calculated SFE is slightly better than the standard inputs, the improvement is minor. As a result, adding thermodynamic calculations does not greatly benefit the ML model.

Finally, the GB algorithm with optimized hyper-parameters was re-trained with 75% data and tested with untouched 25% data. According to Fig. 4(a), almost all the data in the train dataset align with the black dash line, which represents the prediction is the same as experiments. This implies that the ML model successfully correlates the composition and temperature to SFE for the train dataset. This model also performs well on the untouched test dataset, except for one outlier. Based on the inset plot, it is clear that more than 70% of the prediction error is less than 10 mJ/m$^2$, which is close to the experimental error [14,48]. The error in SFE inferred from experiments may come from the equipment used for characterization, the inherent variation of SFE within a sample [36,97], and inaccurate elastic constants used when deducing SFE from experimental observation [7,19]. The performance of ML, thermodynamic, and empirical [14] models on the test dataset of SFE are compared, and the results are shown in Fig. 4(b). It is clear that the MAE and RMSE of the ML model are the smallest amongst the empirical and thermodynamic models, confirming that ML is the most capable model for SFE prediction.

The advantage of the ML model is not only higher accuracy, but also its ability to continuously evolve with more data [98,99]. Once the SFE measurements are reported, the data can be incorporated in the current dataset to gain an improved prediction. Another pathway is to refine the thermodynamic models by improving the low-temperature database using the new lattice stability [86,100,101] coupled with an accurate prediction of interfacial energy to generate more reliable physical model predicted data for ML model training.

## 4. Conclusions

In summary, we evaluated the thermodynamic model for SFE prediction, analyzed the influence of alloying elements on SFE, and built an accurate SFE model using ML. We found that there is a need to significantly improve the low-temperature CALPHAD databases and incorporate better magnetic and interfacial energy models to the thermodynamic model. According to literature and Spearman's correlation analysis, Ni and Fe have a moderate monotonous relationship on SFE for most steels. Meanwhile, elements like C and Mn show complex effects on SFE, which are usually related to the alloying element content and the host composition. The GB algorithm performs best among the thermodynamic and empirical models, which proves the advantages of applying ensembled methods for SFE prediction. Based on these results, we envisage that the data-driven method can accelerate alloy design through optimization of SFE to yield preferred deformation mechanisms and excellent mechanical properties.

**Acknowledgments**

This work was supported by the National Science Foundation [1808082]; University of Pittsburgh [Central Research Development Fund]. APC charges for this article were fully paid by the University Library System, University of Pittsburgh

**Disclosure statement**

The authors declare no competing interests.

**Data availability**

The data and the code that support the results and discussion within this paper are available from the corresponding author upon reasonable request

**Tables and Figures**

Table 1. Descriptive statistics of the database used in this work *

|  | Temperature | C | Cr | Mn | Mo | N | Ni | Si | Al | P | S | Fe | SFE |
|---|---|---|---|---|---|---|---|---|---|---|---|---|---|
| Mean | 289.94 | 0.09 | 15.36 | 5.09 | 0.41 | 0.07 | 11.23 | 0.34 | 0.10 | 0.00 | 0.00 | 67.31 | 30.60 |
| Standard Deviation | 46.87 | 0.26 | 6.39 | 8.18 | 0.89 | 0.15 | 6.87 | 0.99 | 0.52 | 0.01 | 0.00 | 7.19 | 13.55 |
| Min | 94.30 | 0.00 | 0.00 | 0.00 | 0.00 | 0.00 | 0.00 | 0.00 | 0.00 | 0.00 | 0.00 | 47.04 | 3.26 |
| Max | 598.15 | 3.21 | 30.00 | 32.69 | 2.70 | 1.00 | 31.16 | 6.22 | 4.80 | 0.07 | 0.04 | 86.46 | 72.97 |

* Temperature unit is Kelvin, composition are given as wt.%, SFE is in mJ/m$^2$

Table 2. MAE of TCFE9 and TCHEA3 prediction in the different temperature ranges

| Temperature range | MAE of TCFE9 prediction (mJ/m$^2$) | MAE of TCHEA3 prediction (mJ/m$^2$) |
|---|---|---|
| 300 K < T < 600 K | 65.5 | 18.4 |
| T = 300 K | 62.8 | 72.0 |
| 94 K < T < 300 K | 77.3 | 55.4 |

Table 3. Comparison of the performance of different algorithms for different input sets using 10-fold cross-validation (CV)

| Machine learning algorithm | Mean RMSE (mJ/m$^2$) of 10-fold CV | | | Mean MAE (mJ/m$^2$) of 10-fold CV | | |
|---|---|---|---|---|---|---|
| | Standard | With TCFE9 | With TCHEA3 | Standard | With TCFE9 | With TCHEA3 |
| Gradient boosting | 7.8 | 7.8 | 7.9 | 5.5 | 5.5 | 5.5 |
| Random forest | 7.8 | 7.9 | 8.1 | 5.7 | 5.7 | 5.8 |
| XGBoost | 8.1 | 8.0 | 8.2 | 5.6 | 5.7 | 5.9 |
| Huber | 9.3 | 9.2 | 9.4 | 6.9 | 6.9 | 7.1 |
| Adaptive boosting | 9.3 | 9.3 | 9.5 | 7.6 | 7.6 | 7.7 |
| K-nearest neighbors | 9.4 | 8.9 | 9.3 | 6.8 | 6.6 | 7.0 |
| Elastic net | 9.5 | 9.2 | 9.5 | 7.1 | 7.0 | 7.1 |
| Ridge | 9.5 | 9.3 | 9.5 | 7.1 | 7.0 | 7.1 |
| Kernel ridge | 9.5 | 9.3 | 9.5 | 7.1 | 7.0 | 7.1 |
| Lasso | 9.5 | 9.2 | 9.5 | 7.2 | 7.0 | 7.1 |
| Automatic relevance determination | 9.5 | 9.3 | 9.5 | 7.2 | 7.1 | 7.2 |
| LassoLars | 9.5 | 9.3 | 9.5 | 7.2 | 7.1 | 7.2 |
| Extra-trees | 9.5 | 10.5 | 10.2 | 6.9 | 8.0 | 7.2 |
| Monotonic regression | 9.6 | 9.4 | 9.6 | 7.3 | 7.1 | 7.4 |
| Bayesian ridge | 9.6 | 9.3 | 9.5 | 7.2 | 7.0 | 7.1 |
| Multilayer perception | 9.7 | 9.5 | 10.1 | 7.4 | 7.2 | 7.3 |
| Supporting vector machine | 9.8 | 9.3 | 9.7 | 7.0 | 6.8 | 7.0 |
| Decision tree | 9.5 | 10.5 | 9.6 | 7.2 | 6.9 | 7.5 |
| Stochastic gradient descent | 11.4 | 11.3 | 11.3 | 9.3 | 8.9 | 9.1 |

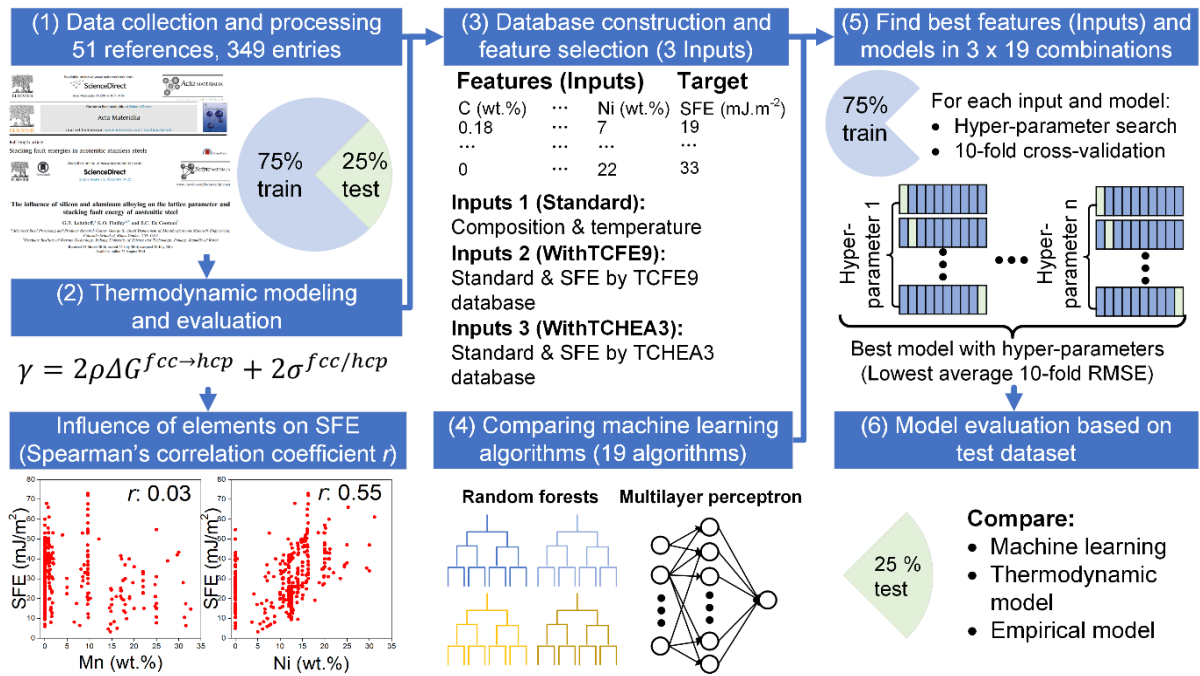

Figure 1. Schematic flow chart of this work, including (1) data collection and curation (2) thermodynamic modeling of SFE (3) database construction and feature selection for machine learning (4) machine learning using 19 algorithms (5) finding best features (inputs) and models (6) model evaluation based on the test dataset.

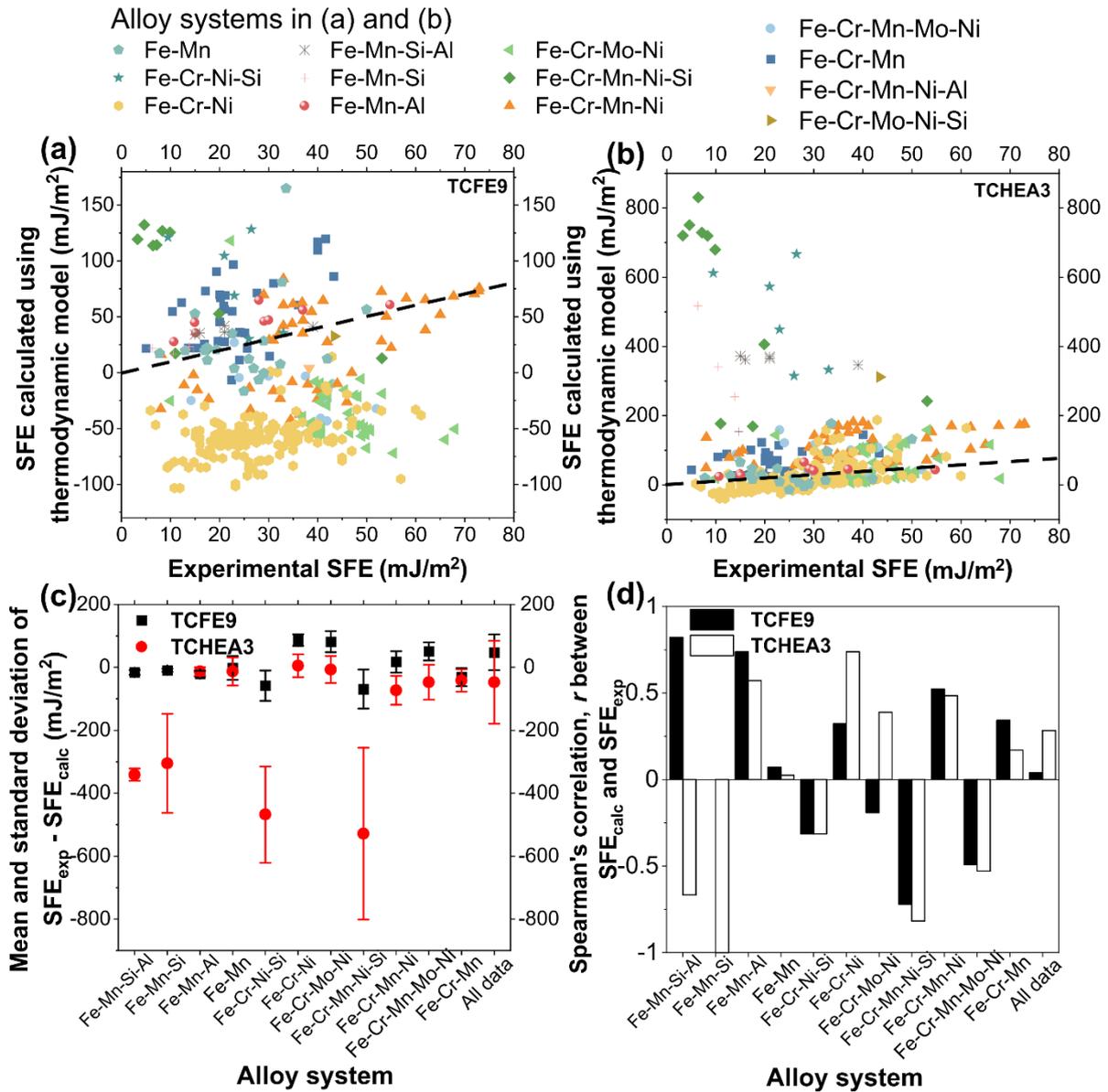

Figure 2. Comparison of SFE for different alloy systems between experimental value ($SFE_{exp}$) and model-prediction based on CALPHAD databases ($SFE_{calc}$): (a) TCFE9 and (b) TCHEA3. The black dashed line indicates the equivalent relationship between $SFE_{calc}$ and $SFE_{exp}$, i.e., $SFE_{calc} = SFE_{exp}$; (c) Mean value and standard deviation for the difference between $SFE_{calc}$ and $SFE_{exp}$ of different alloy systems; (d) The Spearman's correlation coefficient, $r$, between the $SFE_{calc}$ and $SFE_{exp}$ for each alloy system.

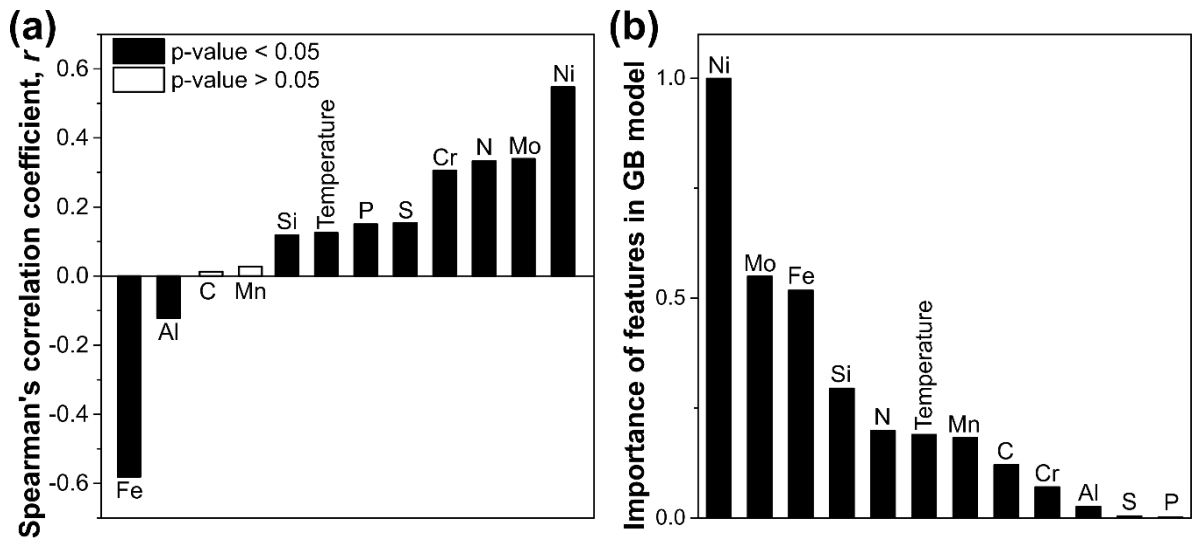

Figure 3. (a) Spearman's correlation coefficient and p-value for the SFE and all features used in this work; (b) Importance of each feature in gradient boosting (GB) model.

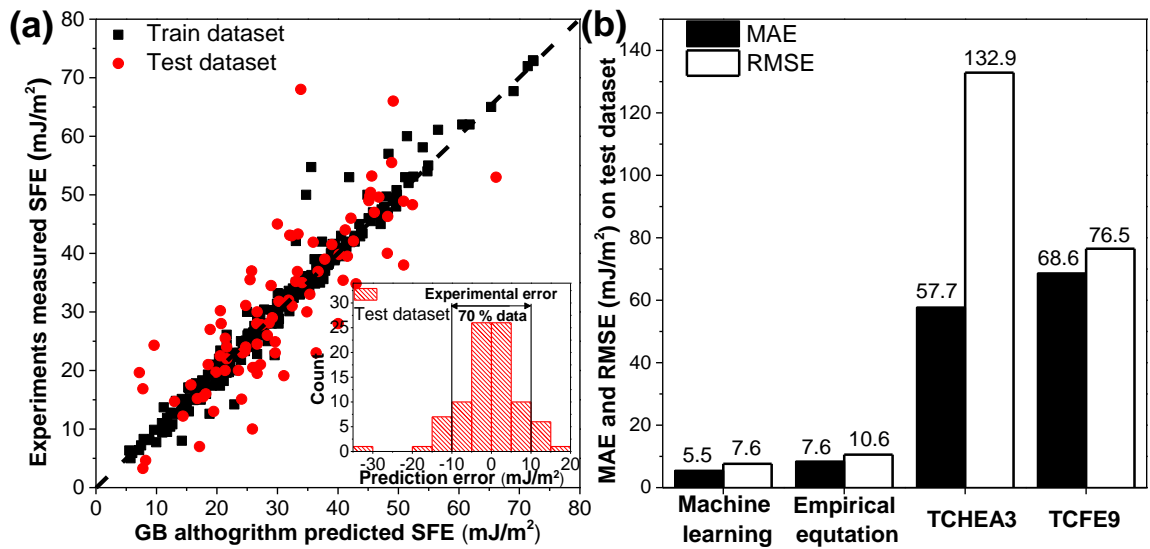

Figure 4. (a) Comparison of the experimental and ML predicted SFE, the black dash line represents the ideal case where prediction equals to measured SFE; (b) Comparison of the model accuracy between machine learning, empirical model, and thermodynamic modeling based on TCHEA3 and TCFE9 databases in terms of MAE and RMSE.